\documentstyle{basi}
\begin{document}
\title[Solar cycle variation in helioseismic data]{ Solar cycle variation in GONG and MDI data: 1995-2002} 
\author[S.C. Tripathy]%
       {S. C. Tripathy \\
        Department of Physics, Indian Institute of Science, Bangalore 560 012\thanks{Permanent address: Udaipur 
Solar Observatory, PRL, PO Box No. 198, Udaipur 313 001}}
\maketitle
\label{firstpage}
\begin{abstract}
Both GONG and MDI projects have measured {\it p}-mode frequencies of the Sun for more than 7 years. 
Here we review what we have learnt from the temporal variation of 
the oscillation frequencies and splitting coefficients.  
\end{abstract}

\section{Introduction}
\label{sec:intro}
Helioseismology has now been accepted as a powerful diagnostic tool to carry information from inside
the sun at different depths. The basic data which enables to disseminiate this information 
are the global oscillation modes which are characterised by the eigenfrequencies. During the solar cycle 21, 
it was observed that mode frequencies change 
 with the evolution of the solar cycle (Woodard and Noyce, 1985; Libbrecht and 
Woodard, 1989; Elsworth et al., 1990) which suggest that these changes are manifestation of 
the variation of the magnetic field
with solar cycle. During the early stages of the current solar cycle 23, 
the frequency variation has also been shown to be strongly 
correlated with many other activity indices at the solar surface ({\it e.g.} Bhatnagar, Jain and Tripathy,  
 1999). 
With nearly 7 years of co-temporal helioseismic data from two different instruments from two different
platforms, we are now in a stronger position to study the time variation of
{\it p}-mode frequencies with changes in the levels of solar activity. 

The solar oscillations are described as spherical waves which sense the spherical geometry of the sun. The 
angular frequencies are represented by spherical harmonics $Y_m^\ell$, where the degree $\ell$ is  the total number 
of nodal lines on the spherical surface while $m$ is the number of nodes along the equator and is restricted
between $-\ell \le m \le \ell$; each mode with a given $\ell$ has $2\ell + 1$ values of $m$ associated with it. 
The cyclic frequencies $\omega_{n\ell m}$ depends on the mode which is lebeled by the value of 
$\ell, m$, and $n$; $n$ being the radial order which signifies the number of nodes in the radial direction 
({\it cf.} Christensen-Dalsgaard, 1998). The frequencies of modes of 
oscillations are typically represented as a mean frequency $\nu_{n\ell}$ ($\omega_{n \ell m}$/$2\pi$) 
and frequency 
splittings $\nu_{n\ell m}$ $-$ $\nu_{n\ell}$ between modes in the same multiplet. 
The frequency splittings of the order $m$ are normally parametrised  
according to the formula
\begin{equation}
\nu_{n\ell m} = \nu_{n\ell} + \sum_{j=1}^{j=j_{max}} a_j (n, \ell)\; p_j^\ell
\end{equation}
where the basis functions are polynomials related to the 
Clebsch-Gordon coefficients (Ritzwoller and Lavely, 1991). The coefficients $a_j$ are normally referred 
as the $a$ coefficients. The odd $a$ coefficients express the difference between prograde 
and retrograde mode frequencies caused by the 
solar rotation and  can be used to calculate the solar rotation. 
The even $a$ coefficients sense longitudinally symmetric but spherically asymmetric 
properties--local variation in the sound speed 
or the asphericity or the magnetic field strength or a combination of the two.

\section{Analysis and results}
\label{sec:data}
For the present study, we consider 68 overlapping GONG data sets (GONG month 2-69)
derived from 108 day long time series but spaced in an interval of 36 days. These data sets   
cover a period from   May 7, 1995 to March 30, 2002 and  were produced through the standard GONG pipeline 
(Hill et al., 1996). Each set yields about
 60,000 useful frequencies for individual $n$, $\ell$, $m$ modes for $\ell$ = 0 to 150 in about 
1600 multiplets. Central frequencies $\nu_{n\ell}$ and $a$ coefficients up to $a_9$ are derived from these 
multiplets.  The MDI data consists of 31 non-overlapping data sets starting at May 1, 1996 and ending 
on November 1,  2002 with  interuptions between June 16 and October 22, 1998 
due to the loss of contact with SOHO. All these data sets are  72 days long except those 
immediately before and after the break, which are shorter.  
Each data set consist of centroid {\it p}-mode frequencies up to $\ell$~$\approx 200$ and 
$a$-coefficients up to $a_{18}$ (Schou, 1999).

\subsection{Frequency shifts}
The mean shift $\delta \nu$ for a given $\ell$ and $n$
is calculated from the relation 
\begin{equation}
\delta\nu(t)  =  {\sum_{n,\ell}\frac{\delta\nu_{n,\ell}(t)}{\sigma_{n,\ell}^2}}/{\sum_{n.\ell}
\frac{1}{\sigma_{n,\ell}^2}} ,
\end{equation}    
where $\delta\nu_{n,\ell}(t)$ is the change in measured frequency for degree $\ell$ and 
radial order $n$ and 
$\sigma_{n,\ell}$ is the error in the observed frequency. Since the solar activity 
changes over the cycle, we have defined the change $\delta \nu_{n,\ell}(t)$ = $\nu_{n,\ell}(t)$ $-$ $<\nu_{n,\ell}>$ 
where the average is taken over all the available data sets instead of a particular reference set. 

\input epsf
\begin{figure}
\centering
 \epsfxsize=5cm \epsfbox{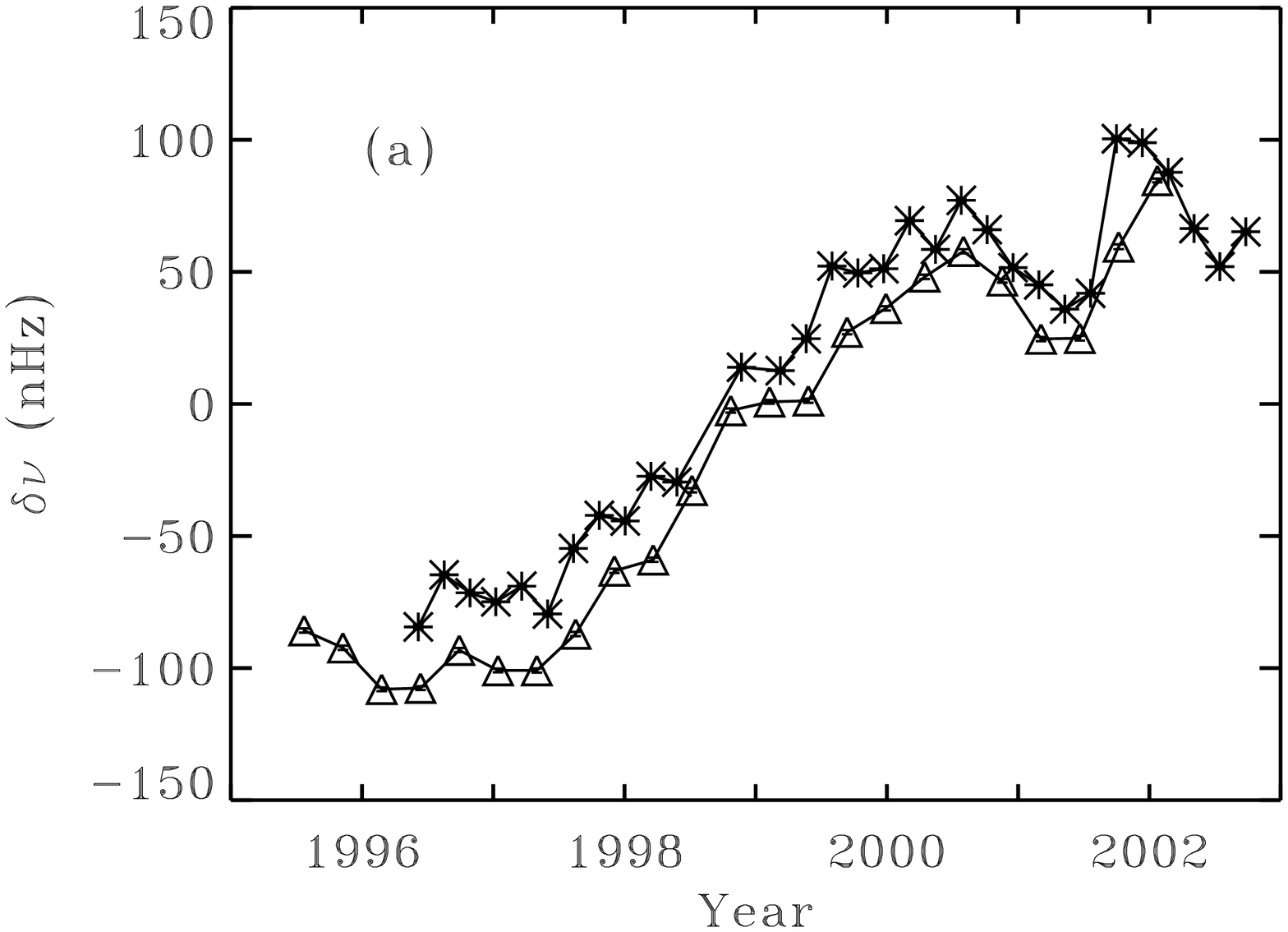}
\epsfxsize=5cm \epsfbox{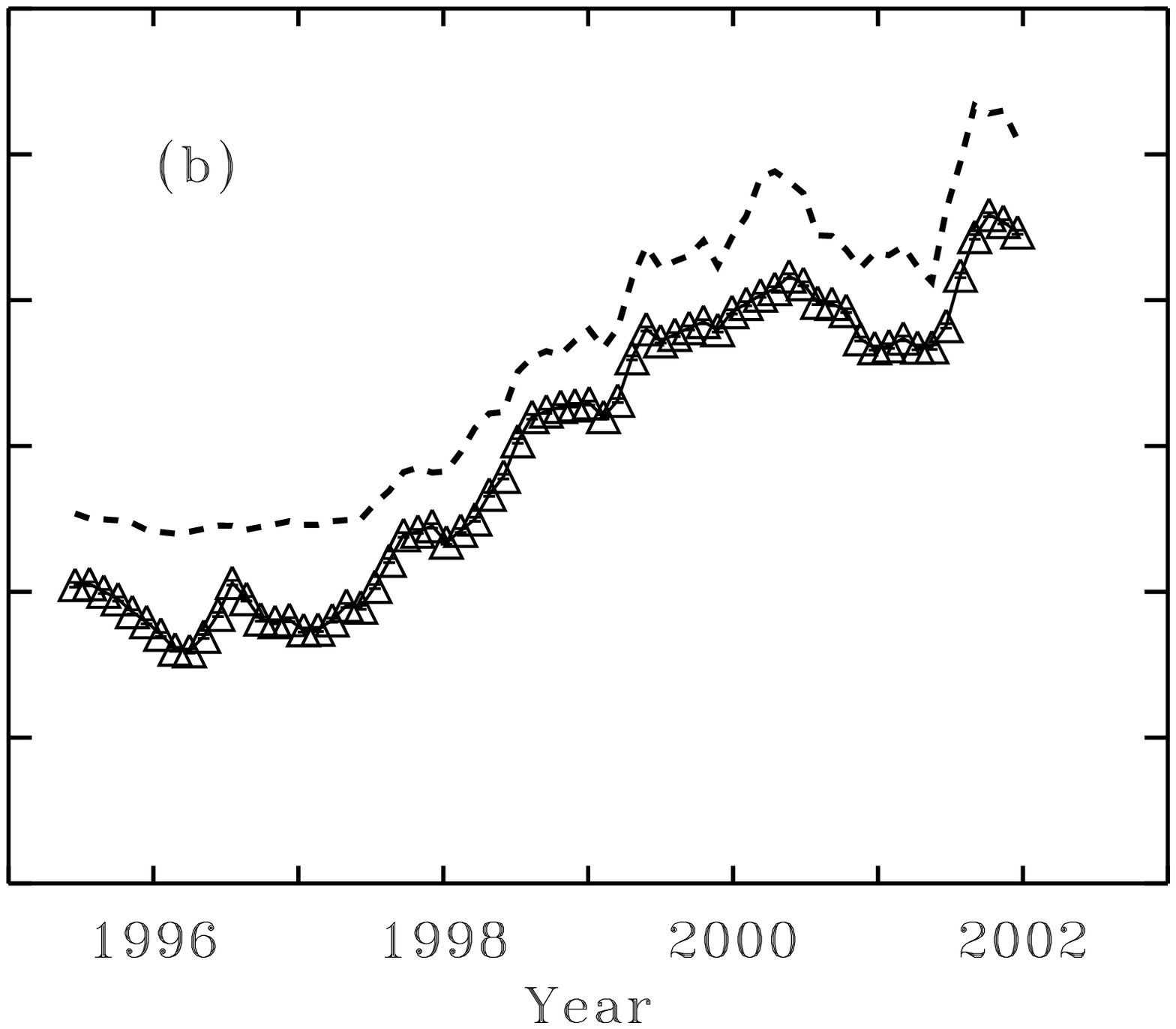}
\caption{The change in mean frequency  for the period 1995-2002. In panel (a), the
triangles represent the shifts calculated from non-overlapping GONG data sets
while the stars represent the shifts from MDI data. Panel (b) shows the 
frequency shifts of the overlapping GONG data sets along with the 
scaled activity level represented by 10.7 cm radio flux (dashed line).} 
\label{fg1}
\end{figure}

The frequency shift is well known to have a strong dependence on frequency and for a meaningful analysis,  
we have considered only those {\it p}-modes which are present in all 
GONG and MDI data sets 
in the frequency range of 1500--3500 $\mu$Hz. 
Figure~1(a) shows the temporal variation of mean frequency shift for MDI and non-overlapping GONG data 
sets  and we clearly 
note that there is a systematic offset between the two shifts. This may have been caused either due to  
different analysis technique to derive the mode frequencies (Schou et al., 2002) or 
because of the different time series lengths over which avearges are taken to compute 
the frequency shifts (also see Jain and Bhatnagar, 2003). The mean frequency shift is correlated with various activity 
indices representing photospheric, 
chromospheric and coronal activities. In panel (b), we show the frequency shifts calculated from  the 
overlapping GONG data sets along with the scaled activity index represented by 10.7 cm radio flux. 
It is evident that the change in frequency follows the 
change in solar activity very closely, the proximity being marginally higher 
during the ascending phase of the  solar cycle. A similar result for the low-$\ell$ {\it p}-mode 
frequencies were reported by Chaplin et al. (2001).  
For a detailed investigation, we calculate the mean frequency shifts corresponding to four different 
frequency bins of 500 $\mu$Hz (Figure~2). The offset between GONG and MDI data sets is quite 
apparent in lower frequency bands and slowly decreases for higher frequency ranges. For the highest
frequency band (3000-3500 $\mu$Hz), the offset is seen only  when the activity level is high.  
In low frequency bands, deviations from the simple activity dependence is observed
during the activity minimum period.  
\begin{figure}[t]
\centering
\epsfxsize=8cm \epsfbox{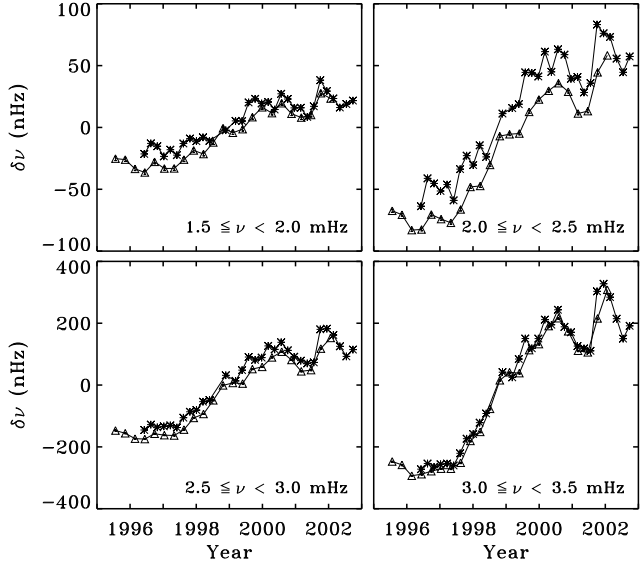}
\vspace{0.5cm}
\caption{Variation of the mean frequency shifts for four different frequency ranges in bins of 500 $\mu$Hz 
for non-overlapping GONG (traingles) and MDI (star) data sets.}
\label{fg2}
\end{figure}

Since both GONG and MDI frequencies are obtained from time series spanning over a long period, 
these can be used to study the temporal evolution of a single ($n$,$\ell$) multiplet. The  variation of 
the central frequency of two multiplets for $n$ = 6 and $n$ = 9 corresponding to $\ell$ = 60,  
along with the scaled activity index 
is shown in Figure 3. It is remarkable that even the frequency of  a single mode of oscillation 
closely follows  the changes in activity level. As mentioned earlier, a 
 small offset between the absolute values for GONG and MDI frequencies at lower $n$ values 
corresponding to lower frequency range is clearly seen. However,  the sensitivity to the 
activity level appears independent of the data sets used. 

\begin{figure}[t]
\centering
\epsfxsize=5cm \epsfbox{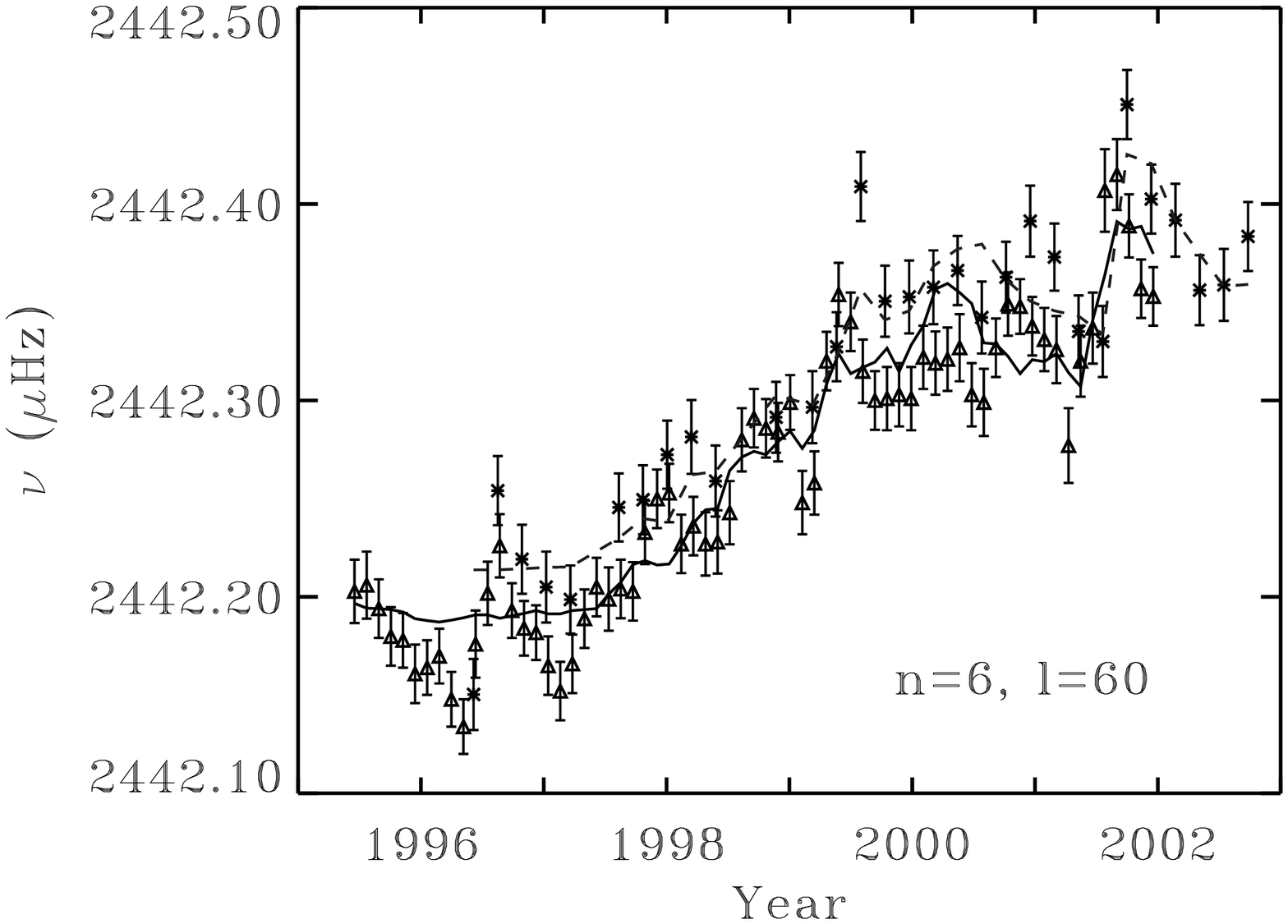}
\hskip 0.5cm
\epsfxsize=5cm \epsfbox{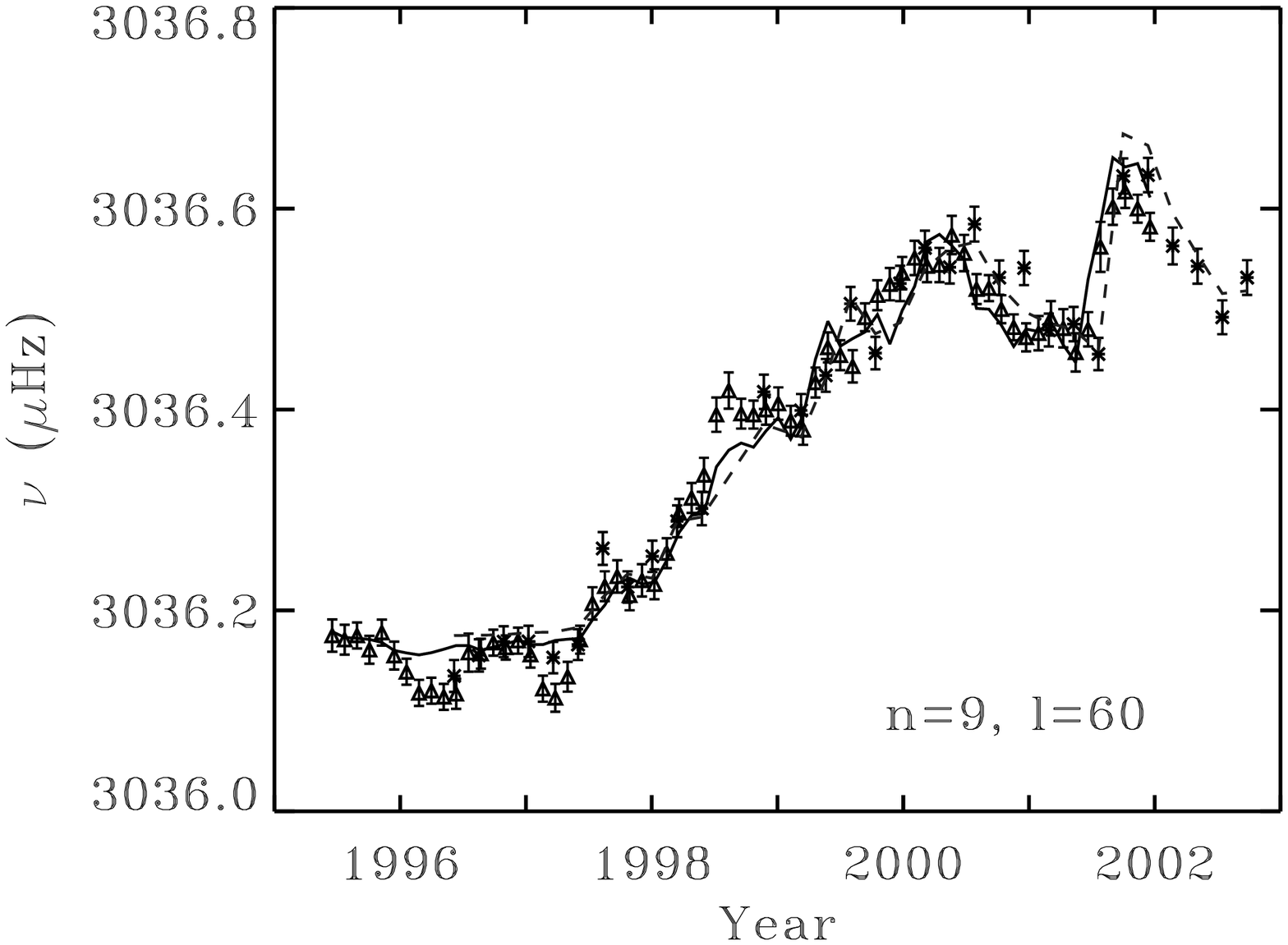}
 \caption{Temporal variation of centroid frequency for  two different values of $n$ 
corresponding to same value of $\ell$ for GONG (triangles) and MDI (stars) data sets.  
 The 1 $\sigma$ errors are also shown.  
 The continuous curves represent 10.7~cm radio 
flux, scaled by the best fit to GONG (solid line) and MDI (dashed line) data. 
}\label{fg3}
\end{figure}

\subsection{Variation in solar rotation rate}
Early helioseismic results have confirmed that the surface differential rotation detected with Doppler
Observations  persists throughout the convection zone (e.g. Brown et al. 1989). 
The rotation rate in the convection zone along different latitudes is nearly constant, while at the base 
of the convection zone, a shear layer (the tachocline) separates the radiative interior which 
rotates almost like a solid body. In solar dynamo theories, it is generally believed that 
the rotation stretches the poloidal field lines near the tachocline and creates the toroidal fields, 
hence it is important to look for possible
 variations in rotation rate with time. There are two appraoches, the first one is an analytical 
approach (Morrow, 1988; Jain, Tripathy and Bhatnagar, 2000) which calculates the surface rotation rates at different 
latitudes by combination of the
odd $a$ coefficients. In the second approach, one uses the inversion techniques which provides 
information both with latitude and depth.  

\begin{figure}[t]
\centering
\epsfxsize=5cm \epsfbox{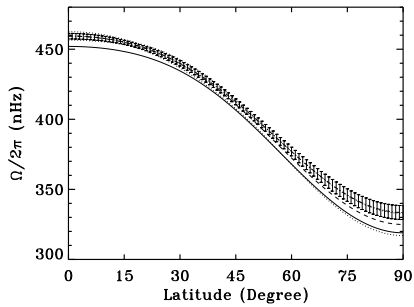}
\caption{Variation of average solar rotation rate as a function of latitude as derived from different data sets. 
The solid line represents the Doppler measurements 
from Snodgrass (1984), the dashed line represenst the non-overlapping GONG data and the dash-dot line represent 
the MDI data.
The 1$\sigma$ uncertainties are shown by the dotted lines for the GONG data and as error bars for the
MDI data} 
\label{fg5}
\end{figure}

Figure~4 shows the rotation rate as a function of latitude from different measurements and clearly displays the 
differential nature of the solar rotation, the maximum rate being at the equator and the minimum at the poles. 
It also shows that the rotation rate derived from the linear combination of the averaged GONG and MDI 
odd $a$'s are in good agreement
with each other and also with those from surface Doppler measurements (Snodgrass, 1984). 
The residual in rotation rate contains the temporally varying 
component of the rotation and are best studied through inversion techniques. 
Inversions show that at low latitudes the bands of faster and slower regions (zonal
flows) move towards the equator with time (Schou, 1999; Antia and Basu, 2000). 
These are analogous to the 
torsional oscillations seen at the solar surface (Howard and LaBonte, 1980) which are believed 
to arise from the nonlinear interactions between magnetic fields and differential rotation. At high latitudes,
the bands seem to move toward the poles (Antia and Basu 2001; Ulrich 2001). The changing pattern of the zonal
flows implies that the maximum and minimum velocities for each latitude occur at different times (Antia and Basu, 2003).  
However, its implication for solar dynamo theory is not well understood. 

\section{Conclusions}
We have analysed the {\it p}-mode oscillation frequencies obtained from GONG and MDI instruments covering a
period of seven  years which includes the descending phase of solar cycle 22 and ascending phase of cycle 23. 
The frequencies show an increase with solar activity and changes are found to be well correlated with 
activity indices. There appears to be a small offset between the frequency shifts derived from the GONG and MDI 
data sets.
A marginally higher correlation is seen for the higher frequencies in the ascending phase of the solar cycle. 
We also note that the residual rotation rate behaves differently at low and high latitudes. 

\section*{Acknowledgements}
I wish to express my thanks to my collaborators Drs. A. Bhatnagar and Kiran Jain.  This work utilises data obtained by 
the Global Oscillation Network Group project, managed by the
National Solar Observatory, which is operated by AURA, Inc., under a cooperative agreement with the National
Science Foundation. The data were acquired by instruments operated by the Big Bear Solar Observatory, 
High Altitude Observatory, Learmonth Solar Observatory, Udaipur Solar Observatory, Instituto de Astrof\'{i}sico
de Canarias, and Cerro Tololo Inter-American Observatory. This work also utilises data from the Solar 
Oscillations Investigation/Michelcon Doppler Imager 
on the Solar and Heliospheric Observatory
(SOHO). SOHO is a project of international cooperation between ESA and NASA.

\label{lastpage}
\end{document}